\documentclass[a4paper,10pt]{article}

\usepackage{graphicx}

\usepackage{amssymb}

 \def\bld#1{{\bf #1 }}

\def\part#1#2{\frac{\partial #1}{\partial #2}}
\def\pder#1#2{{\partial #1/\partial #2}}

\def\rb{\right)}
\def\lb{\left(}

\def\pder#1#2{{\partial #1/\partial #2}}

\title{Statistical mechanics of gravitating systems: An Overview}
\author{T. Padmanabhan\\
Inter-University Centre for Astronomy and Astrophysics,\\
 Post Bag 4, Ganeshkhind, \\
Pune-411 007, India.\\
email: {nabhan@iucaa.ernet.in} }

\date{ }

\begin{document}

\maketitle

\begin{abstract}
I review  several issues related to statistical description of
gravitating systems  in both static and expanding backgrounds.
 After briefly reviewing
the results for the static background, I concentrate on gravitational clustering\index{gravitational clustering}  of collision-less particles in an expanding universe. In particular, I describe 
(a) how the non linear mode-mode coupling transfers power from one scale to another in the Fourier space if the initial power spectrum is sharply peaked at a given scale and
(b) what are the asymptotic characteristics  of gravitational clustering
that are independent of the initial conditions.  Numerical simulations as well as analytic work shows that  power transfer leads to a universal power spectrum at late times, somewhat reminiscent of the existence of Kolmogorov spectrum in fluid turbulence.
\end{abstract}

\section{Overview of the key issues and results}\label{intro}

The statistical mechanics of systems dominated by gravity  has close connections
  with areas of condensed matter physics, fluid mechanics, re-normalization group etc. and
  poses an incredible challenge as regards the basic formulation.  The ideas also find application in many different areas of astrophysics and cosmology,
  especially in the study of globular clusters, galaxies and gravitational clustering in the
  expanding universe. (For a  overall review of statistical mechanics of gravitating systems,
  see \cite{tppr}, \cite{textone}, \cite{texttwo}; review
   of gravitational clustering in expanding background is available in \cite{tpiran} and
  in several textbooks in cosmology \cite{cosmotext};  for a sample of different
  attempts to understand these phenomena by different groups; see \cite{chavanis}, \cite{sanchez},
  \cite{vala}, \cite{fola}, \cite{botta}, \cite{roman}  and the references cited therein.)
  It will be useful to begin with a broad overview
  and a  description of the issues which will be addressed in this article.
  
  In Newtonian theory, the gravitational force can be described
as a gradient of a scalar potential and the evolution of a set of particles under the action of gravitational
 forces can be described the equations
\begin{equation}
 \ddot{\bf x}_i = - \nabla \phi ({\bf x}_i, t); \quad
\nabla^2 \phi = 4 \pi G \sum_i m_i \delta_D ({\bf x} -{\bf x}_i) 
\end{equation}
where  ${\bf x}_i$ is the position of the $i-$th particle, $m_i$ is its
mass.  For an isolated system with 
sufficiently large number of particles, it is useful to investigate whether 
some kind of statistical description  of such a system is possible. 
Such a description, however,  is complicated by  the  
 long range, \index{long range} unscreened, nature of 
 gravitational force.  
 If a 
self gravitating system is divided into two parts, the total energy 
of the system cannot be expressed as the sum of the gravitational
energy of the components. The 
conventional results in statistical physics are
  based on the extensivity of the 
energy \index{extensivity of the 
energy} which is clearly invalid for gravitating systems. 
  To construct the statistical description of such a system, one must begin
  with the construction of the micro-canonical ensemble describing such a system.
  If the Hamiltonian of the system is $H(p_i, q_i)$ then the volume $g(E)$ of the constant
  energy surface  $H(p_i, q_i) =E$ will be of primary importance in the  micro-canonical ensemble\index{micro-canonical ensemble}.
  The logarithm of this function will give the entropy $S(E) = \ln g(E) $ and the temperature of the
  system will be $T(E)\equiv \beta(E)^{-1} = (\pder{S}{E})^{-1}$. 
  
  Systems for which a description based on 
  canonical ensemble is possible, the Laplace transform of $g(E)$ with respect to a variable
  $\beta$ will give the partition function $Z(\beta)$. It is, however, trivial to show that 
  gravitating systems of interest in astrophysics cannot be described by a canonical
  ensemble \cite{tppr}, \cite{dlbone}, \cite{dlbtwo}.
  Virial theorem holds for such systems  and we have $(2K+U) =0$ where 
  $K$ and $U$ are the total kinetic and potential energies of the system.
  This leads to $E=K+U= -K$; since the temperature of the system is proportional
  to the total kinetic energy, the  specific heat\index{specific heat} will be negative: 
  $C_V \equiv (\pder{E}{T})_V 
  \propto (\pder{E}{K}) < 0$.   On the other hand, the specific heat of any system
  described by a canonical ensemble $C_V = \beta^2 <(\Delta E)^2>$ will be
  positive definite. Thus one cannot describe self gravitating systems
  of the kind we are interested in by  canonical ensemble\index{canonical ensemble}.
  
  One may attempt to find the equilibrium configuration for self gravitating
  systems by maximizing the entropy $S(E)$ or the phase volume 
  $g(E)$. It is again easy to show that no global maximum for the entropy
  exists for classical point particles interacting via Newtonian gravity.
  To prove this, we only need to construct a configuration with   arbitrarily high
   entropy which can be achieved as follows:
  Consider a system of $N$ particles initially occupying a region of finite 
  volume in phase space and total energy $E$. We now move  a small
  number of these particles (in fact, a pair of them, say, particles 1 and 2  will do) arbitrarily close to 
  each other. The potential energy of interaction of  these two particles, 
  $-Gm_1m_2/r_{12}$, will become arbitrarily high as $r_{12}\to 0$. Transferring some
  of this energy to the rest of the particles, we can increase their kinetic energy without limit.
  This will clearly increase the phase volume occupied by the system without bound.
  This argument can be made more formal by dividing the original system into a small,
  compact core and a large diffuse halo and allowing the core to collapse
  and transfer the energy to   the halo.

  The absence of the global maximum for entropy --- as 
  argued above --- depends on the idealization
   that there is no short distance cut-off
  in the interaction of the particles, so that we could take the limit $r_{12}\to 0$.
  If we assume, instead, that each particle has a minimum radius $a$, then
  the typical lower bound to the gravitational potential energy contributed by a 
  pair of particles will be $-Gm_1m_2/2a$. This will put an upper bound on the 
  amount of energy that can be made available to the rest of the system.
  
   We have also assumed that part of the system can expand without limit --- in the sense
  that any particle with sufficiently large energy can move to arbitrarily large
  distances. In real life, no system is completely isolated and eventually one has to 
  assume that the meandering particle is better treated as a member of 
  another system. One way of obtaining a truly isolated system is to confine
  the system  inside a spherical region of radius $R$ with, say, reflecting
  wall. (Most of our discussion is confined to 3-dimensions and the situation is diffrent in 2-dimensions; see e.g \cite{two-d})
  
 The two cut-offs $a$ and $R$ will make the upper bound on the entropy finite, but
 even with  the two cut-offs, the primary nature
  of gravitational instability cannot be avoided. The basic phenomenon
  described  above  (namely, the formation of a  compact
  core\index{compact
  core} and a  diffuse halo\index{diffuse halo}) will still occur since this is the direction of increasing
  entropy. Particles in the hot diffuse component will permeate the entire spherical
  cavity, bouncing off the walls and having a kinetic energy which is 
  significantly larger than the potential energy. The compact core will exist as 
  a gravitationally bound system with very little kinetic energy. 
  A more formal way of understanding this phenomena is based 
  on  the virial theorem for a system with a short distance cut-off
  confined to a sphere of volume $V$. In this case, the virial theorem will
  read as \cite{textone} 
  \begin{equation}
  2T + U = 3PV + \Phi
  \label{modvirial}
  \end{equation}
  where $P$ is the pressure on the walls and $\Phi$ is the correction to the potential
  energy arising from the short distance cut-off. This equation can be satisfied
  in essentially three different ways. If $T$ and $U$ are significantly higher than
  $3PV$, then we have $2T + U \approx 0$ which describes a self gravitating 
  systems in standard virial equilibrium but not in the state of maximum entropy.
  If $T \gg U$ and $3PV\gg \Phi$, one can have $2T \approx 3PV$ which
  describes an ideal gas with no potential energy confined to a container of
  volume $V$; this will describe the hot diffuse component at late times.
  If  $T \ll U$ and $3PV\ll \Phi$, then one can have $U\approx \Phi$ describing
  the compact potential energy dominated core at late times.
  In general, the evolution of the system will lead to the production of the core and
  the halo and each component will satisfy the virial theorem in the form (\ref{modvirial}).
  Such an asymptotic state with two distinct phases  \cite{aaron}  is quite different from what would have
  been expected for systems with only short range interaction.
  Considering its importance, I shall briefly describe in section \ref{sec:phases} 
   a toy model which captures the essential
  physics of the above system in an exactly solvable context. 
  
  The above discussion focussed on the existence of global maximum
  to the entropy and we proved that it does not exist in the absence of 
  two cut-offs. It is, however, possible to have {\it local} extrema of entropy 
  which are not global maxima. Intuitively, one would have expected
  the distribution of matter in  the configuration which is a local
  extrema of entropy to be described by a Boltzmann distribution,
  with the density given by $\rho ({\bf x}) \propto \exp[-\beta \phi({\bf x})]$
  where $\phi$ is the gravitational potential related to $\rho$ by Poisson equation.
  This is indeed true; for a formal proof see \cite{tppr}.  This
  configuration is usually called the  isothermal sphere\index{isothermal sphere} (because it can be 
  shown that, among all solutions to this equation, the one with spherical symmetry
  maximizes the entropy) and  it is a local maximum of entropy.  The second (functional)  derivative
  of the entropy with respect to the configuration variables will determine whether
  the local extremum of entropy is a local maximum or a saddle point \cite{antonov}, \cite{tpapjs}.
   
  The relevance of the long range of gravity in all the above phenomena can be
   understood by studying model
  systems with an attractive potential varying as $r^{-\alpha}$ with different values
  for $\alpha$. Such studies confirm the results and interpretation given above;
  (see \cite{ispo} and references cited therein).
  
  Let us now consider the situation in the context of an expanding background.
  There is considerable amount of observational evidence to suggest that
  one of the dominant energy densities in the universe is contributed by self gravitating
  point particles. The smooth average energy density of these particles drive
  the expansion of the universe while any small deviation from the homogeneous energy
  density will cluster gravitationally. [For a review of cosmology from a contemperorary perspective, see e.g., \cite{cosmoreview}] One of the central problems in cosmology is to describe
  the non linear phases of this gravitational clustering starting from a initial spectrum of density
  fluctuations. It is often enough (and necessary) to use a statistical description and  relate
  different statistical indicators (like the  power spectra\index{power spectra}, $n$th order  correlation functions\index{correlation functions} ....)
  of the resulting density distribution to the statistical parameters (usually the power spectrum) of the 
  initial distribution. The relevant scales at which gravitational clustering is non linear are less than
  about 10 Mpc (where 1 Mpc = $3\times 10^{24}$ cm is the typical separation between galaxies in the
  universe) while the expansion of the universe has a characteristic scale of about few thousand 
  Mpc. Hence, non linear gravitational clustering in an expanding universe can 
  be adequately described by Newtonian gravity provided the rescaling of lengths due to the
  background expansion
  is taken into account. This is easily done by introducing a  {\it proper}
   coordinate\index{proper
   coordinate} for the $i-$th particle ${\bf r}_i$,
  related to the  {\it comoving} coordinate\index{comoving coordinate}  ${\bf x}_i$, by ${\bf r}_i = a(t) {\bf x}_i$ with 
  $a(t)$ describing the stretching of length scales due to cosmic expansion. The Newtonian 
  dynamics works with the proper coordinates ${\bf r}_i$ which can be translated
  to the behaviour of the comoving coordinate ${\bf x}_i$ by this rescaling. [This implies that, for all practical purposes, we are still in the domain of Newtonian gravity. There is a far deeper connection between thermodynamics and gravity \cite{grav-thermo} in the general relativistic domain which we will not discuss in these lectures.]

  As to be expected, cosmological expansion  completely changes the nature of the problem because of 
  several new factors which come in:
  (a) The problem has now become time dependent and it will be pointless to look for
  equilibrium solutions in the conventional sense of the word. 
  (b) On the other hand, the expansion of the universe has a civilizing influence on the 
  particles and  acts counter to the tendency of gravity
  to make systems unstable. 
  (c) In any small local region of the universe, one would assume that the conclusions
  describing a finite gravitating system will still hold true approximately. In that case, particles in any small
  sub region will be driven towards configurations of local extrema of entropy (say, isothermal
  spheres) and towards global maxima of entropy (say, core-halo configurations).
  
  An extra feature comes into play as regards the expanding halo from any sub region.
  The expansion of the universe acts as a damping term in the equations of motion
  and drains the particles of their kinetic energy --- which is essentially the lowering of
  temperature of any system participating in cosmic expansion.  This, in turn, 
  helps gravitational clustering since the potential wells of nearby sub regions
  can capture particles in the expanding halo of one region when the kinetic energy of 
  the expanding halo  has been sufficiently reduced. 
  
  The actual behaviour of the system will, of course, depend on the form of $a(t)$.
  However, for understanding the nature of clustering, one can take $a(t) \propto 
  t^{2/3}$ which describes a matter dominated universe with critical density.
  Such a power law has the advantage that there is no intrinsic scale in the problem.
  Since Newtonian gravitational force is also scale free, one would expect some 
  scaling relations to exist in the pattern of gravitational clustering. Converting this intuitive idea into a concrete mathematical statement
  turns out to be non trivial and difficult.

  It is clear that cosmological expansion  introduces  
  several new factors into the problem when compared with the study of statistical mechanics of isolated
  gravitating systems. (For a general review of statistical mechanics of gravitating systems,
  see \cite{tppr}. For a sample of different approaches, see \cite{sample} and the references cited therein. 
  Review of gravitational clustering in expanding background is also available in several textbooks in cosmology \cite{cosmotext,lssu}.)
  Though this problem can be
tackled in a `practical' manner using high resolution numerical simulations (for a review, see \cite{numsim}),
such an approach hides the physical principles which govern 
the behaviour of the system. To understand the physics, it is necessary to
attack the problem from several directions using analytic and semi analytic
methods. Several such attempts exist in the literature based on Zeldovich(like) approximations \cite{za},
path integral and perturbative techniques \cite{pi}, nonlinear scaling relations \cite{nsr} and many others. In spite of all these it is probably fair to say that we still do not have a clear analytic grasp of this problem, mainly because each of these approximations have different domains of validity and do not arise from a central paradigm.

I propose to attack the problem from a different angle, which has not received much attention in the past. The approach begins from the dynamical equation for the the density contrast in the Fourier space  and casts it as an integro-differential equation. Though this equation is known in the literature  (see, e.g. \cite{lssu}), it has received very little attention  because it is not `closed' mathematically; that is, it involves variables which are not natural to the formalism and thus further progress is difficult. I will, however, argue that
there exists a natural closure condition for this
equation  based on Zeldovich approximation thereby allowing us to write down
 a \textit{closed} integro-differential equation for the gravitational potential
in the Fourier space. 

It turns out that this equation can form the basis for several further investigations some of which are described in ref. \cite{gc1} and in the second reference in \cite{tppr}. Here I will concentrate on just two specific features, centered around the following issues:

\begin{itemize}
\item
   If the initial power spectrum is sharply peaked in a narrow band of wavelengths, how does
  the evolution transfer the power to other scales? In particular, does the non linear evolution in the case
of gravitational interactions lead to a universal power spectrum (like the Kolmogorov spectrum in fluid turbulence)? 
 
  \item
  What is the nature of the time evolution at late stages? Does the gravitational clustering at late stages wipe out the memory of initial conditions and evolve in a universal manner?
\end{itemize} 
Fair amount of progress can be made as regards these questions using the integro-differential equation mentioned above and some of these aspects will be discussed in detail.

 \section{Phases of the self gravitating system}\label{sec:phases}

As described in section \ref{intro}  the statistical mechanics of
finite, self gravitating,
systems have the following characteristic features: 
(a) They exhibit negative specific heat while in virial equilibrium.
(b) They are inherently unstable to the formation of a core-halo structure and global maximum for entropy
does not  exist without cut-offs at short and large distances. 
(c) They can be broadly characterized by two phases --- one of which is compact and dominated
by potential energy while the other is diffuse and behaves more or less like an ideal gas.
The purpose of this section is to describe a simple toy model which exhibits all these 
features and mimics a self gravitating system \cite{tppr}.  

Consider a system with two particles described by a Hamiltonian of the form
\begin{equation}H \left(  {\bf P}, {\bf Q}; {\bf p}, {\bf r}\right) = {{\bf P}^2 \over 2M} + {{\bf p}^2 \over 2 \mu} - {Gm^2 \over r }   \end{equation}
where $(  {\bf Q}, {\bf P}) $ are coordinates and momenta of the center of mass, $(  {\bf r}, {\bf p} )$ are the  relative coordinates and momenta, $M = 2 m$ is the total mass, $\mu = m/2 $ is the reduced mass  and $m$ is the mass of the individual particles. This system may be thought of as consisting of  two particles (each  of mass $m$) interacting via gravity. We shall  assume that the quantity $r$ varies in the interval $(  a, R )$. This is equivalent to assuming that the particles are hard spheres of radius $a/2$ and that the system is confined to a spherical box of radius $R$. We will study the ``statistical mechanics'' of this  simple toy model\index{simple toy model}.

 To do this, we shall start with the volume $g(E)$
of the constant energy surface $H=E$.
Straightforward calculation gives
\begin{equation}
 g(E)=  AR^3\int_a^{r_{\rm max}} r^2 dr \left[E+{Gm^2\over r}\right]^2.\label{pv}
 \end{equation} 
where $A=(64 \pi ^5 m^3/3)$.
The range of integration in (\ref{pv})  should be limited to the region
in which the expression in the square brackets is positive. So we
should use $r_{\rm max}= (Gm^2/|E|)$ if $ (-Gm^2/a)< E < (-Gm^2/R) $, and use
$r_{\rm max}= R$ if $(-Gm^2/R)<E<+\infty$. Since $H\geq (-Gm^2/a)$, we trivially
have $g(E)=0$ for $E<(-Gm^2/a)$. The constant $A$ is unimportant for our
discussions and hence will be omitted from the formulas hereafter.
The integration in (\ref{pv})  gives the following result:
\begin{equation}
 {g(E)\over (Gm^2)^3}=\left\{ 
 \begin{array}{l}
  {R^3\over3}(-E)^{-1} \left( 1+{aE\over Gm^2}\right)^3 ,\quad\qquad
           (-Gm^2/a)< E<(-Gm^2/R)\\
           {   }\\
          {R^3\over3} (-E)^{-1} \left[ \left(1+{RE\over Gm^2}\right)^3
          - \left(1+{aE\over Gm^2}\right)^3 \right],
         (-Gm^2/R )<E< \infty .
 \end{array}
        \right.
         \label{gee}\end{equation}
 This function $g(E)$ is continuous and smooth at $E= (-Gm^2/R)$. We define the entropy $S(E)$  and
the temperature $T(E)$ of
the system by the relations  
\begin{equation} S(E)=\ln g(E);\quad T^{-1}(E)=\beta (E)= {\partial S(E)\over \partial E}.
\label{entro}\end{equation}        
All the interesting thermodynamic properties of the system can be understood
from the $T(E)$ curve.

Consider first the case of  low energies with $(-Gm^2/a)<E<(-Gm^2/R)$.
Using (\ref{gee}) and (\ref{entro})  one can easily obtain $T(E)$ and write it in the
 dimensionless form as
\begin{equation} t(\epsilon)= \left[{3\over 1+\epsilon}
 -{1\over \epsilon}\right]^{-1}\label{temp}\end{equation}
where we have defined $t=(aT/Gm^2)$ and $\epsilon =(aE/Gm^2)$.

This function exhibits the peculiarities characteristic of gravitating systems.
At the lowest energy admissible for our system, which corresponds to $\epsilon
=-1$, the temperature $t$ vanishes. This describes a tightly bound
low temperature phase of the system with negligible random motion. 
 The $t(\epsilon)$
is clearly dominated by the first term of (\ref{temp})   for $\epsilon \simeq -1$. As we increase
the energy of the system, the temperature {\it increases}, which is the normal
behaviour for a system. This trend continues up to
\begin{equation}\epsilon =
\epsilon_1= -{1\over2}(\sqrt3 -1) \simeq -0.36  \end{equation}
at which point the $t(\epsilon)$ curve reaches a maximum and turns around.
As we increase the energy further the temperature {\it decreases}.  The
system {\it exhibits negative specific heat in this range.}

Equation (\ref{temp})  is valid  from the minimum energy $(-Gm^2/a)$
 all the way up to the energy $(-Gm^2/R)$. For realistic systems, $R\gg a$
and hence this range is quite wide. For a small region in this range,
[from $(-Gm^2/a)$ to $(- 0.36 Gm^2/a)$] we have positive specific heat;
for the rest of the region the specific heat is negative. {\it The positive
specific heat region owes its existence to the nonzero short distance
cutoff.} If we set $a=0$, the first term in (\ref{temp})   will vanish; we will
have $t\propto (-\epsilon^{-1})$ and negative specific heat in this entire domain.

For $E\geq (-Gm^2/R)$, we have to use the second expression in ({\ref{gee})  for $g(E)$.
In this case, we get:
\begin{equation} t(\epsilon)= \left[ { 3\left[ (1+\epsilon)^2- {R\over a}(1+ {R\over a}\epsilon)^2\right]
                        \over (1+\epsilon)^3 -(1+{R\over a}\epsilon)^3}
                       -{1\over \epsilon} \right]^{-1} .\label{qtemp}\end{equation}
This function, of course, matches smoothly with (\ref{temp})   at $\epsilon=-(a/R)$.
As we increase the energy, the temperature continues to decrease for
a little while, exhibiting negative specific heat. However, this behaviour
is soon halted at some $\epsilon=\epsilon_2$, say. The $t(\epsilon)$ curve reaches
a minimum at this point, turns around, and starts increasing with increasing
$\epsilon$. We thus enter another (high-temperature) phase with positive
specific heat. From (\ref{qtemp})   it is clear that $t\simeq (1/2)
 \epsilon$ for large $\epsilon$. 
(Since $E=(3/2)NkT$ for an ideal gas, we might have expected to find
$t\simeq (1/3)\epsilon$ for our system with $N=2$ at high temperatures.
This is indeed what we would have found if we had defined our entropy as
the logarithm of the volume of the phase space with
$H \le E$. With our definition, the energy of the ideal gas is actually
$E=[{(3/2)}N-1]kT;$ hence we get $t=(1/2)\epsilon$ when $N=2$). 
The form of the $t(\epsilon)$ for $(a/R) = 10^{-4}$  is shown
in figure \ref{fig:phases} by the dashed curve.  The specific heat is positive along the portions AB and CD
and is negative along BC.

     \begin{figure}[ht]
   \begin{center}
   \includegraphics[width=0.85\textwidth]{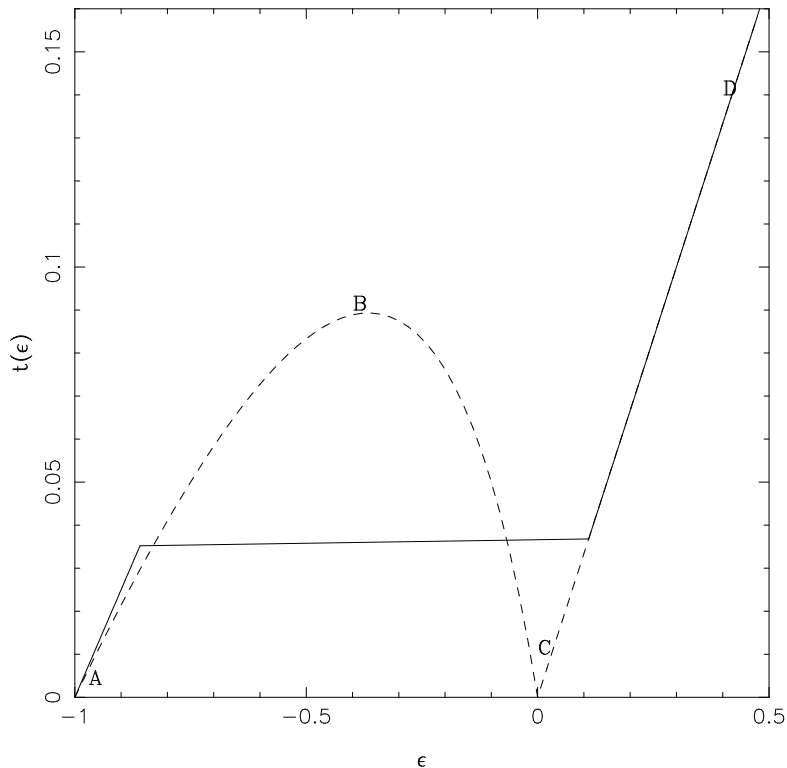}
   \end{center}
   \caption{The relation between temperature and energy for a model mimicking self
   gravitating systems. The dashed line is the result for micro-canonical
   ensemble and the solid line is for canonical ensemble. 
   The negative specific heat region, BC, in the micro-canonical 
   description is replaced by a phase transition in the canonical
   description. See text for more details   }
   \label{fig:phases}
  \end{figure}

The overall picture is now clear. Our system has two natural energy scales:
$E_1=(-Gm^2/a)$ and $E_2=(-Gm^2/R)$. For $E\gg E_2$, gravity is not strong
enough to keep $r<R$ and the system behaves like a gas confined
by the container; we have a  high temperature phase\index{high temperature phase} with positive specific
heat. As we lower the energy to $E\simeq E_2$,
 the effects of gravity begin to be felt.
For $ E_1<E<E_2$, the system is unaffected by either the box or the short
distance cutoff; this is the domain dominated entirely by gravity and
we have negative specific heat. As we go to $E\simeq E_1$,
the hard core nature of the particles begins to be felt and the gravity is
again resisted. This gives rise to a  low temperature phase\index{low temperature phase} with positive
specific heat.

 We can also consider the effect of increasing $R$, keeping
$a$ and $E$ fixed. 
Since we imagine the particles to be hard spheres of radius $(a/2)$,
we should only consider $R>2a$. It is amusing to note that, if
$2<(R/a)<(\sqrt 3 +1)$, there is no region of negative specific heat.
As we increase $R$, this negative specific heat region appears
 and increasing $R$  increases
the range over which the specific heat is negative. Suppose a
system is originally prepared with some $E$ and $R$ values such that the specific heat 
is positive. If we now increase $R$,  the system may find itself
in a region of negative specific heat.{\it This suggests the possibility that
an instability may be triggered in a constant energy system if its radius
increases beyond a critical value.}  We will see later that this is indeed true.

Since systems described
by canonical distribution cannot 
exhibit negative specific heat, it follows that canonical distribution will
lead to a very different physical picture for this range of (mean) energies $E_1<E<E_2$.
It is, therefore, of interest to look at our system from the point of view of
canonical distribution by computing the partition function.
In the partition function
\begin{equation} Z(\beta)=\int d^3Pd^3pd^3Qd^3r \exp (-\beta H)  \end{equation}
 the integrations over $P,p$ and $Q$ can be performed trivially.
Omitting an overall constant which is unimportant, we can write the
answer  in the dimensionless form as
\begin{equation} Z(t)= t^{3} \left({R\over a}\right)^3\int_1^{R/a} dx
 x^2 \exp \left( {1\over xt}\right) \label{parfn}\end{equation}
where $t$ is the dimensionless temperature defined in (\ref{temp})  . Though this
integral cannot be evaluated in closed form, all the limiting properties of
$Z(\beta)$ can be easily obtained from (\ref{parfn}). 

The integrand in (\ref{parfn})  is large for both large and small $x$ and reaches
a minimum for $x=x_m=(1/2t)$.  At high temperatures, $x_m <1$ and hence the minimum falls outside
the domain of integration. The exponential contributes very little to
the integral and we can approximate $Z$ adequately by
\begin{equation}Z\approx t^3\left({R\over a}\right)^3
\int_1^{R/a} dx x^2 \left[1+{2x_{m}\over x}\right]={t^3\over 3}
\left({R\over a}\right)^6\left(1+{3a\over 2Rt}\right)
.\label{hparfn}\end{equation}
On the other hand, if $x_m>1$ the minimum lies between the limits of the
integration and the exponential part of the curve dominates the integral.
We can easily evaluate this contribution by a saddle point approach, and obtain 
\begin{equation}
 Z \approx \left({R\over a}\right)^3t^4 (1-2t)^{-1} \exp\left( {1\over t}\right)  .
\label{lparfn}\end{equation}
As we lower the temperature, making $x_m$ cross $1$ from below, the contribution
switches over from (\ref{hparfn})  to (\ref{lparfn}). The transition is exponentially sharp. The  critical
temperature\index{critical
temperature} at which the transition occurs can be estimated by finding the temperature
at which the two contributions are equal. This occurs at
\begin{equation} t_c={1\over 3}{1\over \ln (R/a)}.\label{tcrit}\end{equation}
For $t<t_c$, we should use (\ref{lparfn})  and for $t>t_c$ we should use (\ref{hparfn}).

Given $Z(\beta)$ all thermodynamic functions can be computed. In particular,
the mean energy of the system is given by
$ E(\beta)= -(\partial \ln Z/ \partial\beta)$.
This relation can be inverted to give the $T(E)$ which can be compared
with the $T(E)$ obtained earlier using the micro-canonical distribution.
 From (\ref{hparfn})  and (\ref{lparfn})  we get,
\begin{equation}\epsilon (t)= {aE\over Gm^2}= 4t-1  \label{lener}\end{equation}
 for $t<t_c$ and
\begin{equation}\epsilon(t)= 3t-{3a\over 2R} \label{hener}\end{equation}
for $t>t_c$. Near $t\approx t_c$, there is a rapid variation of the energy
and we cannot use either asymptotic form. The system undergoes a phase transition
at $t=t_c$ absorbing a large amount of energy
\begin{equation}\Delta \epsilon\approx \left( 1-{1\over 3\ln (R/a)}\right). \end{equation}
The specific heat is, of course, positive throughout the range. This is
to be expected because canonical ensemble cannot lead to negative specific
heats.

The  $T-E$ curves obtained from the canonical (unbroken line)  and micro-canonical (dashed line)
distributions are shown in figure \ref{fig:phases}. 
 (For convenience, we have rescaled the $T-E$ curve of the micro-canonical distribution so that $\epsilon \simeq 3t$ asymptotically.)
At both very low and very high temperatures,   the  canonical and micro-canonical 
descriptions
match. The crucial difference occurs at the intermediate energies and
temperatures. Micro-canonical description predicts negative specific heat and
a reasonably slow variation of energy with temperature. Canonical
description, on the other hand, predicts a phase transition with
rapid variation of energy with temperature. Such phase transitions are
accompanied by large fluctuations in the energy, which is the main reason
for the disagreement between the two descriptions \cite{tppr}, \cite{dlbone}, \cite{dlbtwo}.  

Numerical analysis of more realistic systems confirm all these features. 
Such systems exhibit a phase transition from the  diffuse virialized phase\index{diffuse virialized phase}
to a  core dominated phase\index{core dominated phase} when the temperature is lowered below a critical value \cite{aaron}.
The transition is very sharp and occurs at nearly constant temperature.
The energy released by the formation of the compact core heats up the diffuse
halo component.

\section{Isothermal sphere}\label{isosph}

While a global maximum
  to the entropy  does not exist in the absence of 
  two cut-offs, it is, however, possible to have {\it local} extrema of entropy 
  which are not global maxima. Such a configuration is described a by a Boltzmann distribution,
  with the density given by $\rho ({\bf x}) \propto \exp[-\beta \phi({\bf x})]$
  where $\phi$ is the gravitational potential related to $\rho$ by Poisson equation (for a formal proof see \cite{tppr}).
  Among all solutions to this equation, since the one with spherical symmetry
  maximizes the entropy this
  configuration is usually called the  isothermal sphere.
The extremum condition for the entropy,
 is equivalent to the 
  differential equation
for the gravitational potential:
\begin{equation}\nabla^2 \phi = 4 \pi G \rho_c e^{-{\beta} \left[ \phi \left(  {\bf x} \right)  - \phi \left(  0 \right) 
 \right] } \label{qself} \end{equation}
Given the solution to this equation, all other quantities can be
determined. As we shall see, this system shows
several peculiarities.

It is convenient to introduce the length, mass and energy scale by the definitions
\begin{equation}L_0 \equiv \left(  4 \pi G \rho_c \beta \right) ^{1/2}, \quad M_0 = 4 \pi \rho_c L_0^3, \quad\phi_0 \equiv \beta^{-1} = {GM_0 \over L_0 }   \end{equation}
where $\rho_c = \rho(0)$. All
 other physical variables can be expressed in terms of the 
dimensionless quantities
\begin{equation} x \equiv {r \over L_0}, \quad n \equiv {\rho \over \rho_c }, \quad m= {M \left(  r \right)  \over M_0 }, \quad y \equiv \beta \left[ \phi - \phi \left(  0 \right)  \right] .  \end{equation}
In terms of $y(x)$ the isothermal equation (\ref{qself}) becomes 
\begin{equation}{1\over x^2}{d\over dx}(x^2{dy\over dx})={\rm e}^{-y}\label{qdliso}\end{equation}
with the boundary condition $y(0)=y'(0)=0$. Let us consider the nature of
solutions to this equation.

By direct substitution, we see that $n = \left(  2 /x^2 \right) , m = 2x, y = 2 \ln x $ satisfies these equations. This solution, however, is singular at the origin
and hence is not physically admissible. The importance of this solution
lies in the fact that  other (physically admissible) solutions
tend to this solution \cite{tppr}, \cite{chandra} for large values of $x$. 
This asymptotic behavior of all solutions shows that the density decreases as $(1/r^2)$ for
large $r$ implying that the mass contained inside a sphere of
radius $r$ increases as $M(r)\propto r$ at large $r$. To find physically 
useful solutions, it is necessary to assume that the solution is cutoff at
some radius $R$. For example, one may assume that the system is enclosed
in a spherical box of radius $R$.  In what follows, it will be assumed that
the system has some cutoff radius $R$.

 The equation  (\ref{qdliso})  is invariant under the transformation $y\rightarrow y+a \; ; \;
x\rightarrow kx $ with $k^2= {\rm e}^a$. This invariance implies that,
given a solution with some value of $y(0)$, we can obtain the solution
with any other value of $y(0)$ by simple rescaling. Therefore, only one
of the two integration constants in (\ref{qdliso}) is really non-trivial. 
Hence it must be possible  to reduce the degree of the equation from two to one
by a judicious choice of variables \cite{chandra}. 
One such set of variables are:
\begin{equation}v\equiv{m\over x};\quad u\equiv{nx^3\over m}={nx^2\over v}. \end{equation}
In terms of $v$ and  $u$, equation (\ref{qself}) becomes
\begin{equation}{u\over v}{dv\over du}=-{(u-1)\over (u+v-3)}.\label{quv}\end{equation}
The boundary conditions $y(0) = y'(0)=0$ translate into the
following: $v$ is zero at $u=3$, and $(dv/du)=-5/3$ at (3,0). The solution $v\left(  u \right) $ has to be obtained numerically: it is plotted in figure \ref{figisotherm} 
as the spiraling curve. The singular points of this differential equation are
given by the intersection of the straight lines $u=1$ and $u+v=3$ on which, the
numerator and denominator of the right hand side of (\ref{quv}) vanishes; that is,
the singular point is at $u_s =1$, $ v_s =2$ corresponding to the solution $n = (2/x^2), m=2x$. It is obvious from the 
nature of the equations that the solutions will spiral around the singular
point.

 The nature of the solution shown in figure \ref{figisotherm}  allows us
to put an interesting bounds on physical quantities
including energy.  To see this,
we shall compute the total energy $E$ of the isothermal sphere. The potential
and kinetic energies are
\begin{eqnarray}
U&=&-\int^R_0{GM(r)\over r} {dM\over dr}
dr = - {GM_0^2\over L_0} \int^{x_0}_0 mnxdx \nonumber\\ 
K&=&{3\over 2}{M\over \beta}={3\over 2}{GM^2_0\over L_0}
m(x_0) = {GM^2_0\over L_0}{3\over 2}\int^{x_0}_0
nx^2 dx \end{eqnarray}
where $x_0=R/L_0$. The total energy is, therefore, 
\begin{eqnarray}
E&=&K+U={GM^2_0\over 2L_0}\int^{x_0}_0 dx
(3nx^2-2mnx)\nonumber\\ 
 &=& {GM^2_0\over 2L_0}\int^{x_0}_0
dx{d\over dx}\{2nx^3-3m\}
= {GM^2_0\over L_0}\{n_0 x_0^3-{3\over 2}m_0\}  \end{eqnarray}
where $n_0=n(x_0)$ and $m_0=m(x_0)$. The dimensionless quantity
$(RE/GM^2)$ is given by
\begin{equation}\lambda={RE\over GM^2}={1\over v_0}
\{u_0-{3\over 2}\}. \end{equation}
{\it Note that the combination $(RE/GM^2)$ is a function of
$(u,v)$ alone}. Let us now consider the constraints on $\lambda$. 
Suppose we specify some value for $\lambda$ by specifying $R,E$ and
$M$. Then such an isothermal sphere {\it must} lie on the curve
\begin{equation}v={1\over \lambda}\left(  u-{3\over 2}\right) ; \qquad \lambda \equiv \frac{RE}{GM^2}\label{qeline}\end{equation}
which is a straight line through the point $(1.5,0)$ with the
slope $\lambda^{-1}$. On the other hand, since {\it all} isothermal spheres 
must lie on the $u-v$ curve, {\it an isothermal sphere can exist only
if the line in (\ref{qeline}) intersects the $u-v$ curve}.

\begin{figure}[ht]
\begin{center}
\includegraphics[width=.8\textwidth]{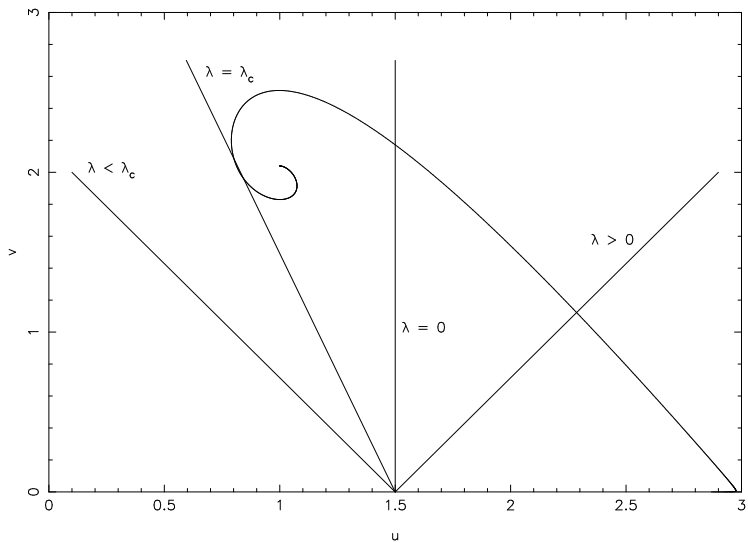}
\end{center}
\caption[]{Bound on $RE/GM^2$ for the isothermal sphere}
\label{figisotherm}
\end{figure}

 For large positive $\lambda$ (positive $E$)
there is just one intersection. When $\lambda=0$, (zero energy) we still have
a unique isothermal sphere. (For $\lambda =0 $, equation (\ref{qeline}) is a vertical line
through $u=3/2$.). When $\lambda$ is negative (negative $E$), the line can cut
the $u-v$ curve at more than one point; thus more than one isothermal 
sphere can exist with a given value of $\lambda$. [Of course, specifying 
$M,R,E$ individually will remove this non-uniqueness]. But as we decrease
$\lambda$ (more and more negative $E$) the line in (\ref{qeline}) will slope more
and more to the left; and when $\lambda$ is smaller than a critical value
$\lambda_c$, the intersection will cease to exist. {\it Thus no isothermal sphere
can exist if $(RE/GM^2)$ is below a  critical value $\lambda_c$.}\footnote{This derivation
is due to the author \cite{tpapjs}.
It is surprising that Chandrasekhar, who
has worked out the isothermal sphere in uv coordinates as early as 1939, missed discovering
the energy bound shown in figure \ref{figisotherm}.
 Chandrasekhar \cite{chandra} has the uv curve but does not 
over-plot lines of constant $\lambda$. If he had done that, he would have discovered  Antonov
instability\index{Antonov
instability} decades before Antonov did \cite{antonov}.}
This fact follows immediately from the nature of $u-v$ curve 
and equation (\ref{qeline}). The value of $\lambda_c$ can be found from the numerical solution in figure. It turns out to be about ($-0.335$). 

The isothermal sphere  has a special status as  a solution to the mean field
 equations. 
Isothermal spheres, however,  cannot exist if $(RE/GM^2) < -0.335$. Even when $(RE/GM^2)>-0.335$, the isothermal solution need not be stable. The stability of this solution can be investigated by studying the second variation of the entropy.
Such a detailed analysis shows that the following results are true \cite{antonov}, \cite{dlbwood},
\cite{tpapjs}.  
(i) Systems with $(RE/GM^2)<-0.335$ cannot evolve into isothermal
spheres. Entropy has no extremum for such systems.
(ii) Systems with ($(RE/GM^2)>-0.335$) and ($\rho(0)> 709\,\rho(R)$) can
exist in a meta-stable (saddle point state) isothermal sphere configuration. Here $\rho(0)$ and $\rho(R)$ denote the densities at the center and edge respectively. The entropy extrema exist but they are not local maxima.
(iii) Systems with ($(RE/GM^2)> -0.335$) and ($\rho(0)<709\,\rho(R)$) can
form isothermal spheres which are local maximum of entropy.

\section{An integral equation to describe nonlinear gravitational clustering}\label{gravclnl}

Let us next consider the gravitational clustering of a system of collision-less point particles 
{\it in an expanding universe} which poses
several challenging theoretical questions. Though the problem can be
tackled in a `practical' manner using high resolution numerical simulations,
such an approach hides the physical principles which govern 
the behaviour of the system. To understand the physics, it is necessary that we
attack the problem from several directions using analytic and semi analytic
methods. These sections will describe such attempts and will emphasize
the semi analytic approach and outstanding issues, rather than more well established results. 

The expansion of the universe sets a natural length scale (called the Hubble radius) $d_H = c
(\dot a/a)^{-1}$ which is about 4000 Mpc in the current universe. 
In any region which is small compared to $d_{\rm H}$ one can set up an unambiguous coordinate system in which the {\it proper} coordinate of a particle ${\bf r} (t)=a(t){\bf x}(t)$ satisfies the Newtonian equation $\ddot {\bf r} = -  {\nabla }_{\bf r}\Phi$ where $\Phi$ is the gravitational potential. 
The Lagrangian for such a system of particles is given by
\begin{equation}
L=\sum_i\left[
\frac{1}{2}m_i\dot{\mathbf{r}}_i^2+\frac{G}{2}\sum_j\frac{m_im_j}{|\mathbf{r}_i-\mathbf{r}_j|}
\right]
\end{equation} 
In the term
\begin{equation}
\frac{1}{2}\dot{\mathbf{r}}_i^2
=\frac{1}{2}\left[a^2\dot{\mathbf{x}}_i^2+\dot{a}^2\mathbf{x}_i^2+a\dot{a}\frac{d\mathbf{x}_i^2}{dt}\right]
=\frac{1}{2}\left[a^2\dot{\mathbf{x}}_i^2-a\ddot{a}\mathbf{x}_i^2 +\frac{da\dot{a}\mathbf{x}_i^2}{dt}\right]
\end{equation} 
we note that: (i) the total time derivative can be ignored; (ii) using $\ddot{a}=-(4\pi G/3)\rho_ba$, the term $\Phi_{FRW}=-(1/2)a\ddot{a}\mathbf{x}_i^2
=(2\pi G / 3)\rho_0 (x_i^2/a) $ can be identified as the gravitational potential due to the uniform Friedman background
of density $\rho_b=\rho_0/a^3$. Hence the Lagrangian can be expressed as
\begin{equation}
L=\sum_i m_i\left[\frac{1}{2}a^2\dot\mathbf{x}_i^2 +\phi(t,\mathbf{x}_i)\right]=T-U
\label{basicL}
\end{equation}
where
\begin{equation}
\phi=-\frac{G}{2a}\sum_{j}\frac{m_j}{|\mathbf{x}_i-\mathbf{x}_j|}-\frac{2\pi G\rho_0}{3}\frac{x_i^2}{a}
\label{defphi}
\end{equation} 
 is the difference between the total potential and the potential for the back ground Friedman universe $\Phi_{FRW}$. Varying the Lagrangian in Eq.(\ref{basicL}) with respect to $\mathbf{x}_i$, we get the equation of motion to be:
\begin{equation} 
\ddot{\bf x} + 2 {\dot a \over a}\dot{\bf x} = - {1 \over a^2} \nabla_x \phi\ 
\label{traj}
\end{equation}
Since $\phi$ is the gravitational potential generated by the \textit{perturbed}  mass density, 
  it satisfies the equation with the source $(\rho-\rho_b)\equiv
 \rho_b\delta$:
\begin{equation}
 \nabla^2_x \phi  = 4 \pi G \rho_ba^2 \delta 
 \label{poisson}
  \end{equation}
 Equation~(\ref{traj}) and Eq.(\ref{poisson}) govern the nonlinear gravitational clustering in an expanding background.

Usually one is interested in the evolution of the density contrast $\delta \lb t, \bld x \rb $ rather than in the trajectories. Since the density contrast can be expressed in terms of the trajectories of the particles, it should be possible to write down a differential equation for $\delta (t, \bld x)$ based on the equations for the trajectories $\bld x (t)$ derived above. It is, however, somewhat easier to write down an equation for $\delta_{\bld k} (t)$ which is the spatial Fourier transform of $\delta (t, \bld x)$. To do this, we begin with the fact that the density $\rho(\bld x,t)$ due to a set of point particles, each of mass $m$, is given by
\begin{equation}
\rho (\bld x,t) = {m\over a^3 (t)} \sum\limits_i \delta_D [ \bld x - \bld x _i(t)]
\end{equation}
where $\bld x_{i}(t)$ is the trajectory of the ith particle and $\delta_D$ is the Dirac delta function.  The density contrast $\delta (\bld x,t)$ is related to $\rho(\bld x, t)$ by
\begin{equation}
1+\delta (\bld x,t) \equiv {\rho(\bld x, t) \over \rho_b} = {V \over N} \sum\limits_i \delta_D [\bld x - \bld x_i(t)] =  \int d {\bld q} \delta_D [\bld x - \bld x_{T} (t, \bld q)]  . 
\end{equation}
In arriving at the last equality we have taken the continuum limit by: (i) replacing $\bld x_i(t)$ by $\bld x_T(t,\bld q)$ where  $\bld q$ stands for a set of parameters (like the initial position, velocity etc.) of a particle; for
simplicity, we shall take  this to be initial position. The subscript `T' is just to remind ourselves that $\bld x_T(t,\bld q)$ is the
\textit{trajectory} of the particle. (ii) replacing $(V/N)$ by $d^3{\bld q}$ since both represent volume per particle. Fourier transforming both sides  we get
\begin{equation}
\delta_{\bld k}(t) \equiv \int d^3\bld x   {\rm e}^{-i\bld k \cdot \bld x} \delta (\bld x,t) =   \int d^3 {\bld q} \  {\rm exp}[ - i {\bf k} . {\bf x}_{T} (t, \bld q)]  -(2 \pi)^3 \delta_D (\bld k)
\end{equation}
Differentiating this expression, 
and using  Eq.~(\ref{traj}) for the trajectories give one can obtain an equation for $\delta_{\mathbf{k}}$
(see e.g. ref.\cite{gc1}; Eq.10).
The structure of this equation can be simplified if we use the perturbed gravitational potential (in Fourier space) $\phi_{\bf k}$ related to $\delta_{\bf k}$ by 
\begin{equation}
\delta_{\bf k} = - {k^2\phi_{\bld k} \over 4 \pi G \rho_b a^2} = - \lb {k^2 a \over 4 \pi G \rho_0}\rb \phi_{\bld k} = - \lb {2 \over 3H_0^2 }\rb k^2a \phi_{\bld k}
\end{equation}
In terms of $\phi_{\bld k}$ the  \textit{exact} evolution equation reads as:
\begin{eqnarray}
\ddot \phi_{\bf k} + 4 {\dot a \over a} \dot\phi_{\bf k}   &= & - {1 \over 2a^2} \int {d^3{\bf p} \over (2 \pi )^3} \phi_{{1\over 2}{\bf k+p}}   
\phi_{{1\over 2}{\bf k-p}}\left[ \left( {k\over 2}\right)^2 + p^2 
-2  \lb {\bld k . \bld p \over k}\rb^2 \right] \nonumber \\
&+ &\lb{3H_0^2 \over 2}\rb  \int {d^3{\bf q} \over a} \lb{\bld k} . \dot {\bld x}\over k\rb ^2 e^{i{\bf k}.{\bf x}} \label{powtransf} 
\label{evphi}
\end{eqnarray}
where $\bld x = \bld x_T(t, \bld q)$. 
Of course, this equation is not `closed'.
 It contains the velocities of the particles $\dot {\bf x}_T$ and their positions explicitly in the second term on the right and one cannot --- in general --- express them in simple form in terms of $\phi_\mathbf{k}$. As a result, it might seem that we are in no better position than when we started. I will now motivate a strategy to tame this term in order to close this equation. This strategy  depends on two features: 
 
 (1) First, extremely nonlinear structures do not contribute to the right hand side of Eq.(\ref{evphi}) though, of course, they contribute individually to the two terms. 
   More precisely, the right hand side of Eq.(\ref{evphi}) will lead to a density contrast that
    will scale as 
    $\delta_{\bf k} \propto k^2$ if originally  --- in linear 
   theory --- $\delta_k \propto k^n$ with $n>2$ as $k\to 0$.  This leads to a $P\propto k^4$ tail in the power spectrum \cite{lssu,gc1}. (We will provide a derivation of this result in the next section.)

  (ii) Second, we can use Zeldovich approximation to evaluate this term, once the above fact is realized.
  It is well known that, when the density contrasts are small, it grows as
   $\delta_{\bf k}\propto a$  in the linear limit. One can easily show that such a growth corresponds to particle displacements of the form

\begin{equation}
 \bld x_{T} (a,{\bf q}) = {\bf q}  - a \nabla \psi(\mathbf{q});\qquad \psi\equiv (4\pi G\rho_0)^{-1}\phi
 \label{trajec}
\end{equation} 
A useful approximation to describe the quasi linear stages of clustering is obtained by using the trajectory in Eq.(\ref{trajec})  as an ansatz valid {\it even at quasi linear epochs}. In this approximation, (called  Zeldovich approximation), the velocities $\dot\mathbf{x}$ can be expressed in terms of the initial gravitational potential.

We now combine the two results mentioned above  to obtain a closure condition
for our dynamical equation. At any given moment of time
we can divide the particles in the system into three different sets. First, there are particles which are 
already a part of virialized cluster in the non linear regime. Second set is made of particles
which are completely unbound and are essentially contributing to power spectrum at the 
linear scales. The third set is made of all particles which cannot be put into either of these
two baskets. Of these three, we know that the first two sets  of particles do not
contribute significantly to the right hand side of Eq.(\ref{evphi}) so we will not incur any serious error in ignoring these particles in computing
the right hand side. For the description of particles 
in the third set, the Zeldovich approximation should be fairly good. In fact, we can do slightly
better than the standard Zeldovich approximation. We note that in Eq.~(\ref{trajec}) the velocities were
taken to be proportional to the gradient of the \textit{initial} gravitational potential.
We can improve on this ansatz by taking the velocities to be given by the gradient of the 
\textit{instantaneous} gravitational potential which has the effect of incorporating the
influence of particles in bound clusters on the rest of the particles to certain extent.

Given this ansatz, it is straightforward to obtain a \textit{closed} integro-differential
equation for the gravitational potential. Direct calculation shows that the gravitational potential is described by the closed integral equation: (The details of this derivation
can be found in ref. \cite{gc1} and will not be repeated here.)
\begin{equation}
\ddot \phi_{\bld k} + 4 {\dot a \over a} \dot \phi_{\bld k} = -{1 \over 3a^2} \int {d^3 \bld p \over (2 \pi)^3} \phi_{{1 \over 2} \bld k + \bld p} \phi_{{1 \over 2} \bld k - \bld p} 
\left[
{7 \over 8} k^2 + {3 \over 2} p^2 - 5 \lb {\bld k \cdot \bld p\over k}\rb^2
\right] 
\label{key1}
\end{equation}
This equation provides a powerful method for analyzing non linear clustering since estimating Eq.(\ref{evphi}) by Zeldovich approximation has a very large domain of applicability. 
 In the next two sections, I will use this equation to study the transfer of power in gravitational clustering.

 \section{Inverse cascade in non linear gravitational clustering: The $k^4$ tail}\label{nltail}

  There is an interesting and curious result which is characteristic of gravitational
  clustering that can be obtained directly from our Eq.~(\ref{key1}). Consider an initial
  power spectrum which has very little power at large scales; more precisely, we shall
  assume that $P(k)\propto k^n$ with $n>4$ for small $k$ (i.e, the power dies faster than $k^4$ for small $k$). If these large spatial scales are described
  by linear theory --- as one would have normally expected --- then the power at these scales 
  can only grow as $P\propto a^2k^n$ and it will always be sub dominant to $k^4$. It turns out that this 
  conclusion is incorrect.   
As the system evolves, small scale nonlinearities will
develop in the system and --- if the large scales have too little
power intrinsically  ---  then
the long wavelength power will soon be dominated by the
``$k^4$-tail'' of the short wavelength power arising from the
nonlinear clustering.  This is a purely non linear effect which we shall now describe.
(This result is known in literature \cite{lssu,gc1} but we will derive it from the formalism developed in the last section which adds fresh insight.)

\begin{figure}[ht]
\begin{center}
\includegraphics[width=.8\textwidth]{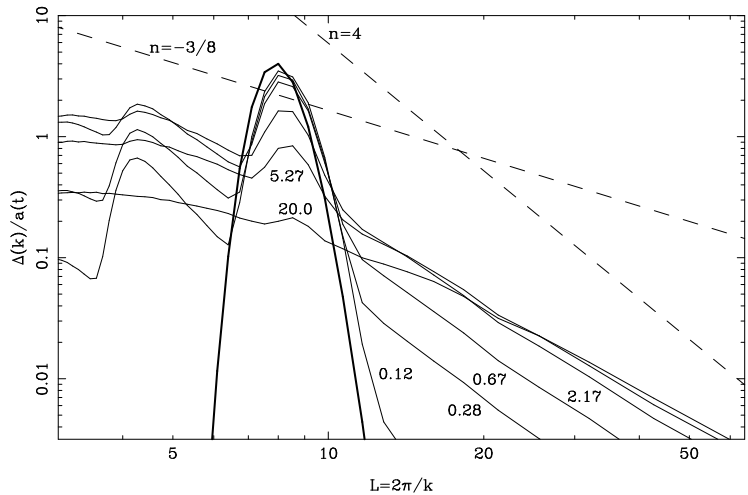}
\end{center}
\caption{The transfer of power to long wavelengths forming a $k^4$ tail is illustrated using 
simulation results. Power is injected in the form of a narrow peak at $L=8$. 
Note that the $y-$axis is $(\Delta/a)$ so that
there will be no change of shape of the power spectrum under linear evolution
with $\Delta\propto a$. As time goes on a $k^4$ tail is generated purely due to nonlinear coupling between the modes. (Figure adapted from ref.\cite{jsbtp1}.)
}
\label{figptsimu}
\end{figure}
  
A  formal way of obtaining the $k^4$ tail is to solve  Eq.~(\ref{key1}) for long wavelengths; i.e. near $\bld k = 0$.   Writing $\phi_{\bld k} = \phi_{\bld k}^{(1)} + \phi_{\bld k}^{(2)} + ....$ where $\phi_{\bld k}^{(1)} = \phi_{\bld k}^{(L)}$ is the time {\it independent} gravitational potential in the linear theory and $\phi_{\bld k}^{(2)}$ is the next order correction, we get from Eq.~(\ref{key1}), the equation  
\begin{equation}
\ddot\phi_{\bld k}^{(2)}+ 4 {\dot a \over a} \dot\phi_{\bld k}^{(2)} \cong - { 1 \over 3a^2} \int {d^3 \bld p \over (2 \pi)^3} \phi^L_{{1 \over 2} \bld k + \bld p} 
\phi^L_{{1 \over 2} \bld k - \bld p} \mathcal{G}(\bld k, \bld p)
\end{equation}
where $\mathcal{G}(\bld k, \bld p)\equiv [(7/8) k^2 + (3/2) p^2 - 5(\bld k \cdot \bld p/k))^2$.
The solution to this equation is the sum of a solution to the homogeneous part [which decays as 
$\dot\phi\propto a^{-4}\propto t^{-8/3}$ giving $\phi\propto t^{-5/3}$] and a particular solution which grows as $a$. Ignoring the decaying mode at late times and taking
$\phi_{\bld k}^{(2)} = aC_{\bld k}$ one can determine $C_{\bld k}$ from the above equation. Plugging it back, we find the lowest order correction to be,
\begin{equation}
\phi_{\bld k}^{(2)} \cong - \lb {2a \over 21H^2_0}\rb \int {d^3 \bld p \over (2 \pi)^3}\phi^L_{{1 \over 2} \bld k + \bld p} 
\phi^L_{{1 \over 2} \bld k - \bld p} \mathcal{G}(\bld k, \bld p)
\label{approsol}
\end{equation}
Near $\bld k \simeq 0$, we have
\begin{eqnarray}
\phi_{\bld k \simeq 0}^{(2)} &\cong& - {2a \over 21H^2_0} \int {d^3 \bld p \over (2 \pi)^3}|\phi^L_{\bld p}|^2 \left[ {7 \over 8}k^2 + {3 \over 2}p^2 - {5(\bld k \cdot \bld p)^2 \over k^2} \right] \nonumber \\
&=&   {a \over 126 \pi^2H_0^2} \int\limits^{\infty}_0 dp  p^4 |\phi^{(L)}_{\bld p}|^2\nonumber \\
\end{eqnarray}
which is independent of $\bld k$ to the lowest order. Correspondingly the power spectrum for
 density $P_{\delta}(k)\propto a^2 k^4P_{\varphi} (k) \propto a^4 k^4$ in this order. 

The generation of long wavelength $k^4$ tail is easily seen in simulations if one starts with a power spectrum that is sharply peaked in $|\bld k|$. Figure \ref{figptsimu} (adapted from \cite{jsbtp1}) shows the results of such a simulation.   The y-axis is $[\Delta(k)/a(t)]$
where $\Delta^2(k) \equiv k^3P/2\pi^2$ is the power per logarithmic band in $k$. 
 In linear theory $\Delta \propto a$ and this quantity should not change. The curves labelled by $a=0.12$ to $a=20.0$ show the \textit{effects of nonlinear evolution}, especially the development of $k^4$ tail.
 (Actually one can do better. The formation can also reproduce the sub-harmonic at $L\simeq 4$ seen in 
 Fig.~ \ref{figptsimu} and other details; see ref. \cite{gc1}.)

\section{Analogue of Kolmogorov spectrum for gravitational clustering}

If 
power is injected at some scale $L$ into an ordinary viscous fluid, it cascades down to smaller scales because of the
non linear coupling between different modes. The resulting power spectrum, for a wide range of scales, is well approximated by the Kolmogorov spectrum which plays a key, useful, role in the study of fluid turbulence. It is possible to obtain the form of this spectrum
from fairly simple and general considerations though the actual equations of fluid turbulence are 
intractably complicated. 
 Let us now consider the corresponding question for non linear gravitational clustering. If power is injected at a given length scale
very early on, how does the dynamical evolution transfers power to other scales at late times?
In particular, does the non linear evolution lead to an analogue of Kolmogorov spectrum with some level of universality, in the case
of gravitational interactions?

Surprisingly, the answer is ``yes", even though normal fluids and collisionless self gravitating particles constitute very different physical systems. If power is injected
at a given scale $L = 2\pi/k_0$ then the gravitational clustering transfers the power to both
larger and smaller spatial scales. At large spatial scales the power spectrum goes as $P(k) \propto k^4$
as soon as non linear coupling becomes important. We have already seen this result in the previous section.  More interestingly, the cascading of power to smaller scales leads to a 
\textit{universal pattern} at late times just as in the case of fluid turbulence. This is because, Eq.(\ref{key1}) admits solutions for the  gravitational potential of the form $\phi_{\bf k}(t) = F(t)D({\bf k}) $ at late times when the initial condition is irrelevant; here
 $F(t)$ satisfies a non linear differential equation and $D({\bf k})$ satisfies an integral
equation. One can  analyze the relevant equations analytically as well as verified the conclusions
by numerical simulations. This study (the details of which can be found in ref.\cite{stp}) confirms that non linear gravitational clustering does
lead to a universal power spectrum at late times if the power is injected at a given scale
initially. 
(In cosmology there is very little  motivation to study the transfer of power by itself
and most of the numerical simulations in the past concentrated on evolving broad band initial power spectrum. So this result was missed out.)

 Our aim is to look for \textit{late time} scale free evolution of the system exploiting the 
fact  Eq.(\ref{key1})
allows self similar solutions of the form 
$\phi_{\bf k}(t) = F(t)D({\bf k}) $. Substituting this ansatz into Eq.~(\ref{key1}) we obtain
two separate equations for $F(t)$ and $D({\bf k})$. It is also convenient at this stage
to use the expansion factor $a(t) = (t/t_0)^{2/3}$ of the matter dominated universe
as the independent variable rather than the cosmic time $t$. Then simple algebra shows that
the governing equations are 
\begin{equation}
a \frac{d^2 F}{da^2} + \frac{7}{2} \frac{dF}{da} = - F^2
\label{tevl}
\end{equation}
and 
\begin{equation}
 H_0^2 D_{\bf k} =\frac{1}{3}\int \frac{d^3{\bf p}}{(2\pi)^3} D_{{1 \over 2} \bld k + \bld p} D_{{1 \over 2} \bld k - \bld p} \mathcal{G} (\bld k, \bld p)
\label{shape}
\end{equation}
Equation~(\ref{tevl}) governs the time evolution while Eq.~(\ref{shape}) governs the shape of
the power spectrum. (The  separation ansatz, of course, has the scaling freedom
$F\to \mu F, D\to (1/\mu)D$ which will change the right hand side of Eq.~(\ref{tevl}) to $-\mu F^2$ and the left hand side of Eq.~(\ref{shape})
to $\mu H_0^2 D_{\bf k}$. But, as to be expected, our results will be independent of $\mu$; so we have set it to unity). Our interest lies in analyzing the solutions of Eq.~(\ref{tevl}) subject to the initial conditions $F(a_i) =$ constant, $(dF/da)_i =0$ at some small enough $a=a_i$.

\begin{figure*}[t]
\begin{center}
\includegraphics[scale=0.5]{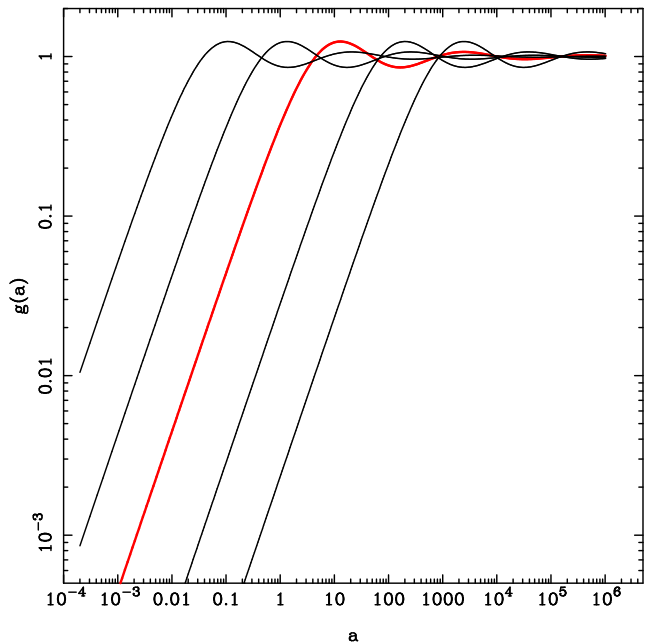}
\end{center}
\caption{The solution to Eq.~(\ref{gofa}) plotted in  $g - a$ plane. The 
function $g(a)$ asymptotically approaches unity with oscillations which are represented by the
spiral in the right panel. The different curves in the  left panel corresponds to the rescaling freedom in the 
initial conditions. One fiducial curve which was used to model the simulation is shown in the
red. For more details, see ref.\cite{stp}.
}
\label{gc1}
\end{figure*}

\begin{figure*}
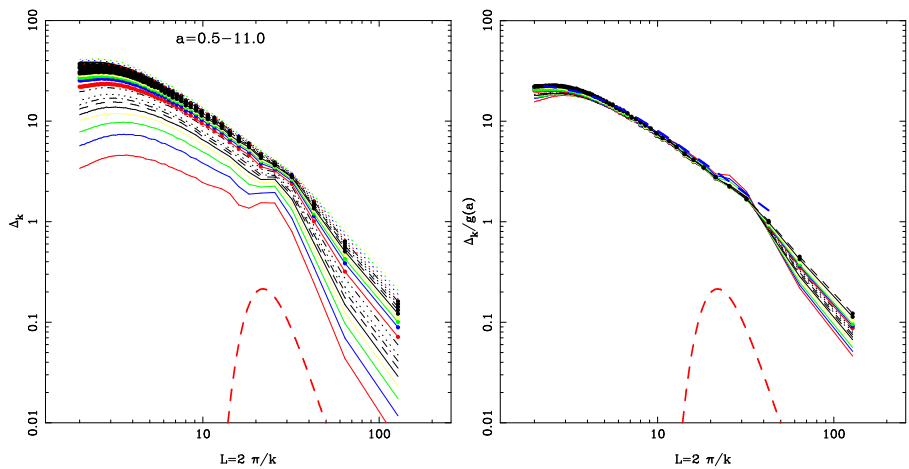

\includegraphics[scale=0.35]{padmanabhan-f3.ps}\ \includegraphics[scale=0.35]{padmanabhan-f4.ps}
\caption{Left panel: The results of the numerical simulation with an initial power spectrum which is 
a Gaussian peaked at $L=24$. The y-axis gives $\Delta_k$ where $\Delta_k^2=k^3 P/2\pi^2$ is
the power per logarithmic band. The evolution generates a well known $k^4$ tail at large scales (see for example, \cite {jsbtp1})
and leads to cascading of power to small scales. Right panel: The simulation data is re-expressed
by factoring out the time evolution function $g(a)$ obtained by integrating Eq.~(\ref{gofa}). The fact
that the curves fall nearly on top of each other shows that the late time evolution is scale free
and described by the ansatz discussed in the text. The rescaled spectrum is very well described by
$P(k)\propto k^{-0.4}/(1+(k/k_0)^{2.6})$ which is shown by the, completely overlapping, broken blue curve.
For more details, see ref.\cite{stp}.
}
\label{fig:gausspeak}
\end{figure*}

Inspection shows that Eq.~(\ref{tevl}) has the exact solution $F(a) = (3/2) a^{-1}$.
This, of course, is a special solution and will not satisfy the relevant initial conditions. However,
Eq.~(\ref{tevl}) fortunately belongs to a  class of non linear equations which can be mapped
to a homologous system. In such cases, the special power law solutions will arise as
the asymptotic limit. (The example well known to astronomers is that of isothermal sphere \cite{chandra}. Our
analysis below has a close parallel.)
To find the general behaviour of the solutions to Eq.~(\ref{tevl}), we will make the substitution
$F(a) = 
(3/2) a^{-1} g(a)$ and change the independent variable from $a$ to $q = \log a$.
Then Eq.~(\ref{tevl}) reduces to the form
 \begin{equation}
\frac{d^2 g}{d q^2} + \frac{1}{2} \frac{d g}{d q} + \frac{3}{2} 
g(g - 1) = 0
\label{gofa}
\end{equation} 
This represents a particle moving in a potential $V(g)= (1/2) g^3 - (3/4) g^2$ under 
friction.  For our initial conditions the motion will lead the ``particle'' to asymptotically come to rest at the stable minimum
at $g=1$ with damped oscillations. In other words, $F(a) \to (3/2) a^{-1}$ for large $a$
showing this is indeed the asymptotic solution. From the Poisson equation, it follows that $k^2\phi_\mathbf{k}\propto (\delta_\mathbf{k}/a)$ so that $\delta_\mathbf{k}(a)\propto g(a)k^2D(\mathbf{k})$ giving a direct physical meaning to the function $g(a)$ as the growth factor for the density contrast. The asymptotic limit ($g\simeq 1$) corresponds to to a rather trivial case of $\delta_\mathbf{k}$ becoming independent of time. What will be more interesting --- and accessible in simulations --- will be the approach to this asymptotic solution. To obtain this, we introduce
the variable 
\begin{equation}
u = g + 2 \left( \frac{dg}{dq}\right)
\end{equation}
so that our system becomes homologous. It can be easily shown that 
we now get the first order form of the autonomous system to be 
\begin{equation} 
\frac{d u}{d g} = - \frac{6 g(g - 1)} {u - g}
\end{equation}
The critical points  of the system are at $(0,0) $ and $(1,1)$. Standard analysis shows that: (i) the  ($0,0$) is an unstable critical point and the second one $(1,1)$ is the stable critical point; (ii) for our initial
conditions the solution spirals around the stable critical point.

Figure \ref{gc1} (from ref. \cite{stp})  describes the solution in the $g-a$  plane. The $g(a)$ curves clearly approach the asymptotic 
value of $g\approx 1$ with superposed oscillations. The different curves in Fig. (\ref{gc1})
are for different initial values which arise from the scaling freedom mentioned earlier. (The thick red line correspond to the initial conditions used in the simulations described below.).  The solution
$g(a)$ describes the time evolution and solves the problem of determining asymptotic time evolution.

To test the correctness of these conclusions, we performed a high resolution simulation 
using the TreePM method
\cite{2002JApA,2003NewA} and its parallel version
\cite{2004astro} with $128^3$ particles
on a $128^3$ grid. 
Details about the code parameters can be found in
\cite{2003NewA}.
The initial power spectrum $P(k)$ was chosen to be a Gaussian peaked at
the scale of $k_p = 2 \pi / L_p$ with $L_p = 24$ grid lengths and with a
standard deviation $\Delta k = 2 \pi / L_{box}$, where $L_{box} = 128$ is
the size of one side of the simulation volume. The amplitude of the peak
was taken such that $\Delta_{lin}\left(k_p = 2 \pi / L_p, a = 0.25\right)
= 1$.
 
The  late time evolution of the power spectrum (in terms of $\Delta_k^2\equiv k^3P(k)/2\pi^2$ where $P=|\delta_k|^2$ is the power spectrum of density fluctuations) obtained from the simulations is shown in Fig.\ref{fig:gausspeak} (left panel). In the right panel, we have rescaled the $\Delta_k$, using the appropriate solution $g(a)$. The fact that the curves fall on top of each other shows that the late time evolution indeed sales as $g(a)$ within numerical accuracy.  A reasonably accurate fit for $g(a)$ at late times used in this figure is given by $
g(a)\propto a(1-0.3\ln a)$.
The key point to note is that the asymptotic time evolution is essentially $\delta(a)\propto a$ except for a logarithmic correction, \textit{even in highly nonlinear scales}. (This was first noticed from somewhat lower resolution simulations in \cite{jsbtp1}.). Since the evolution at \textit{linear} scales is always $\delta\propto a$,
 this allows for a form invariant evolution of power spectrum at all scales. Gravitational clustering evolves towards this asymptotic state. 
 
 To the lowest order of accuracy, the power spectrum at this range of scales is approximated by the mean index $n\approx -1$ with $P(k)\propto k^{-1}$. A better fit to the 
  power spectrum in Fig.\ref{fig:gausspeak} is given by
\begin{equation}
P(k)\propto \frac{k^{-0.4}}{1+(k/k_0)^{2.6}}; \qquad \frac{2\pi}{k_0}\approx 4.5
\label{fit}
\end{equation}
This fit is shown by the broken blue line in the figure which completely overlaps with the data and is barely visible. (Note that this fit is applicable only at $L<L_p$ since the $k^4$ tail will dominate scales to the right of the initial peak; see the discussion in \cite {jsbtp1}). At nonlinear scales $P(k)\propto k^{-3}$
making $\Delta_k$ flat, as seen in Fig. \ref{fig:gausspeak}. (This is \textit{not} a numerical artifact and we have sufficient dynamic range in the simulation to ascertain this.) At quasi linear scales
$P(k)\propto k^{-0.4}$. The effective index of the power spectrum varies between $-3$ and $-0.4$ in this range of scales. 

What could be a possible interpretation for this behaviour ? It is difficult to provide a simple but precise answer but one possible line of reasoning is as follows:
In the case of viscous fluids, the energy is dissipated at the smallest scales as heat. In steady state, energy cannot accumulate at any intermediate scale and hence the rate of flow of energy from one scale to the next (lower) scale must be a constant. This constancy immediately leads to Kolmogorov spectrum. In the case of gravitating particles, there is no dissipation and each scale will evolve towards virial equilibrium. At any given time $t$, the power would have cascaded down only up to some scale $l_{min}(t)$ which it self, of course, is a decreasing function of  time. So, at time $t$ we expect very little power for $1<kl_{min}(t)$ and a $k^4$ tail for $kL_{p}<1$, say. The really interesting band is between $l_{min}$ and $L_{p}$. 

To understand this band, let us recall that the Lagrangian in Eq.(\ref{basicL}) leads to the time dependent Hamiltonian is $H(\mathbf{p},\mathbf{x},t)=\sum[p^2/2ma^2+U]$. The evolution of the energy in the system is governed by the equation $dH/da=(\partial H/\partial a)_{\mathbf{p},\mathbf{x}}.$
It is clear from Eq.(\ref{defphi}) that $(\partial U/\partial a)_{\mathbf{p},\mathbf{x}}=-U/a$ while $(\partial T/\partial a)_{\mathbf{p},\mathbf{x}}=-2T/a$. Hence the time evolution of the total energy $H=E$ of the system is described by
\begin{equation}
\frac{dE}{da}=-\frac{1}{a}(2T+U)=-\frac{2E}{a}-\frac{U}{a}=-\frac{E}{a}-\frac{T}{a}
\end{equation} 
In the continuum limit, ignoring the infinite self-energy term,  the potential energy can be written as:
\begin{equation}
U=-\frac{G\rho_0^2}{2a}\int d^3\mathbf{x}\int d^3\mathbf{y}
\frac{\delta(\mathbf{x},a)\delta(\mathbf{y},a)}{|\mathbf{x}-\mathbf{y}|}
\end{equation} 
Hence
\begin{equation}
\frac{d (a^2E)}{da}=-a^2U=\frac{G\rho_0^2a}{2}\int d^3\mathbf{x}\int d^3\mathbf{y}
\frac{\delta(\mathbf{x},a)\delta(\mathbf{y},a)}{|\mathbf{x}-\mathbf{y}|}
\label{pottot}
\end{equation} 
The ensemble average of the right hand side, per unit proper volume will be
\begin{equation}
\mathcal{E}= -\frac{G\rho_0^2}{2Va^2}\int d^3\mathbf{x} d^3\mathbf{y}
\frac{\langle \delta(\mathbf{x},a)\delta(\mathbf{y},a)\rangle}{|\mathbf{x}-\mathbf{y}|}
\propto \int d^3k \frac{|\delta_k|^2}{a^2k^2} 
\propto\int_0^\infty \frac{dk}{k} \frac{kP(k)}{a^2}
\label{enden}
\end{equation}
where $V$ is the comoving volume.

When a particular scale is virialized, we expect $\mathcal{E}\approx$ constant at that scale in comoving coordinates. That is, we would expect
the energy density in Eq.(\ref{enden}) would have  reached equipartition and contribute same amount per logarithmic band of scales
 in the intermediate scales between $l_{min}$ and $L_{peak}$. 
\textit{ This requires $P(k)\propto a^2/k$ which is essentially what we found from simulations.}
The time dependence of $P$ is essentially $P\propto a$ except for a logarithmic correction. Similarly the scale dependence is $P\propto k^{-1}$ 
 which is indeed a good fit to the simulation results. The flattening of the power at small scales, modeled by the more precise fitting function in Eq.(\ref{fit}), can be understood from the fact that, equipartition is not yet achieved at smaller scales. The same result holds for kinetic energy if the motion is dominated by scale invariant radial flows \cite{jsbtp1,klypin}. Our result suggests that gravitational power transfer evolves towards this equipartition. 
 
 \section*{Acknowledgements}
 
 I thank the organisers of the Les Houches School for inviting me to give these lectures.

\end{document}